\documentclass[aps,prb,amsmath,amssymb,twocolumn,superscriptaddress,showpacs,floatfix]{revtex4-1}

\usepackage{graphicx}
\usepackage{dcolumn}
\usepackage{bm}

\usepackage{amsmath}
\usepackage{amssymb}

\DeclareMathOperator{\me}{\mathit m_{\mathrm e}}

\DeclareMathOperator{\Jsd}{\mathit J_{\mathrm{sd}}}

\DeclareMathOperator{\eps0}{\varepsilon_0}
\DeclareMathOperator{\epsi}{\varepsilon}
\DeclareMathOperator{\klp}{\mathit k^+_\mathrm l}
\DeclareMathOperator{\krp}{\mathit k^+_\mathrm r}
\DeclareMathOperator{\klm}{\mathit k^-_\mathrm l}
\DeclareMathOperator{\krm}{\mathit k^-_\mathrm r}

\begin{document}

\title{Influence of image forces on the interlayer exchange interaction in  magnetic tunnel junctions with ferroelectric barrier}

\author{O.~G.~Udalov}
\affiliation{Department of Physics and Astronomy, California State University Northridge, Northridge, CA 91330, USA}
\affiliation{Institute for Physics of Microstructures, Russian Academy of Science, Nizhny Novgorod, 603950, Russia}
\author{I.~S.~Beloborodov}
\affiliation{Department of Physics and Astronomy, California State University Northridge, Northridge, CA 91330, USA}

\date{\today}

\pacs{75.50.Tt 75.75.Lf	75.30.Et 75.75.-c}

\begin{abstract}
We study interlayer exchange interaction in magnetic tunnel junctions 
with ferroelectric barrier. We focus on the influence of image forces on the voltage dependence of the interlayer magnetic interaction (magneto-electric effect). The influence of the image forces is twofold: 1) they significantly enforce magneto-electric effect 
occurring due to the surface charges at the interface between ferroelectric and ferromagnets; 
2) in combination with voltage dependent dielectric constant of the ferroelectric barrier image forces 
cause an additional contribution to the magneto-electric effect in magnetic tunnel junctions. This contribution can 
exceed the one coming from surface charges. We compare the interlayer exchange coupling 
voltage variation with spin transfer torque effect and show that for half-metallic 
electrodes the interlayer exchange coupling variation is dominant and defines the 
magnetic state and dynamics of magnetization in the tunnel junction. 
\end{abstract}

\maketitle
\section{Introduction}

\begin{figure}
	\includegraphics[width=1\columnwidth]{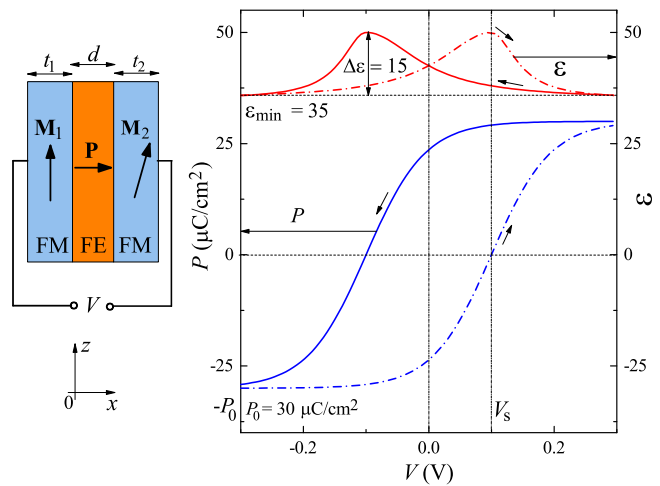}
	\caption{(Color online) Left panel. Magnetic tunnel junction with ferroelectric barrier biased with voltage $V$. $\mathbf M_{1,2}$ are the magnetizations of left and right leads, $\mathbf P$ is the ferroelectric polarization, $d$ is the spacer thickness, $t_{1,2}$ are the electrodes thicknesses. Right panel. Red lines are the dependencies of the dielectric constant $\epsi$ of ferroelectric layer on voltage. Two curves correspond to two polarization states. Black arrows shows hysteresis loop bypass. 
	$V_\mathrm s$ is the switching voltage. $\epsi_{\mathrm{min}}$ and $\Delta\epsi$ are determined in Eq.~(\ref{Eq:Diel}). 
	Blue lines demonstrate polarization of ferroelectric layer as a function of voltage. 
	$P_0$ is the saturation polarization. 
	Shown curves correspond to ferroelectric Hf$_{0.5}$Zr$_{0.5}$O$_2$. }\label{Fig:Model}
\end{figure}

Interlayer exchange coupling (IEC) in magnetic tunnel junction (MTJ) is the long standing problem in the field of spintronics~[\onlinecite{Slonczewski1989,Steren2016,Lee2009,SUZUKI2008,
Ralph2008,Schuhl2002,Bruno1995,Lesnik2006,Dieny2007,Slonczewski2005,
Tsymbal2006,Shvets2009,Blugel2005,Baberschke2003}]. IEC is the interaction between magnetic moments of MTJ electrodes. It is 
described as the surface energy term of the form $E_\mathrm{ex}=-J(\mathbf M_1\cdot\mathbf M_2)/(|\mathbf M_1||\mathbf M_2|)$, where $\mathbf M_{1,2}$ are the magnetizations of ferromagnetic (FM) leads (see Fig.~\ref{Fig:Model}) 
and $J$ is the coupling constant. The coupling induces an effective magnetic field acting on FM layers  in MTJ, $H^i_\mathrm{ex}=J/(|\mathbf M_i|t_i)$, with $t_i$ being the thickness of $i$-th layer. This field may exceed coercive fields of the magnetic leads~[\onlinecite{Tsymbal2006,Lesnik2006,Schuhl2002}]. Thus, the IEC effect is of crucial importance for MTJ magnetic state. Moreover, it is shown theoretically and experimentally that IEC effect can be controlled with voltage ($V$) applied to MTJ~[\onlinecite{Brataas2008,Car2009,Dieny2007,Dijken2013,Lee2009,SUZUKI2008,Ralph2008}]. This opens an avenue 
to the voltage-based magnetization switching in MTJ, 
which is the crucial issue for magnetic memory applications. 
Note that IEC voltage dependence can be considered as magneto-electric (ME) effect~[\onlinecite{Scott2006}].

Applying voltage to MTJ results in a charge current which is spin 
polarized due to FM nature of electrodes. Such a current causes the so-called spin transfer torque (STT) effect~[\onlinecite{Slonczewski1989,Zangwill2002}], which is actively studied at now~[\onlinecite{Brataas2008,Car2009,Dieny2007,Dijken2013,Lee2009,SUZUKI2008,Ralph2008}]. STT effect leads to dynamics of magnetization and can be used to control MTJ magnetic state. However, switching of magnetization in MTJ with STT 
effect appears only at huge currents overheating the system. Another issue is the dynamical nature 
of STT based magnetization switching requiring sophisticated tuning of voltage pulse. 

The IEC effect is not caused by electrical current 
flowing across the MTJ and exists even at zero voltage. 
In contrast, the STT effect is directly related to the electron flow 
between electrodes of the tunnel junction (TJ) and may not be associated 
with any energy term. This is a fundamental difference between IEC and STT effects. 
Spin transfer torque enters the 
macroscopic equation for magnetization dynamics as an additional dissipation term of the form 
$\dot{\mathbf m}_{1,2}=(\gamma J_\mathrm d/(|\mathbf M_{1,2}| t_{1,2}))[\mathbf m_{1,2}\times[\mathbf m_{1,2}\times \mathbf m_{2,1}]]$, where $J_\mathrm d$ is the strength of the STT effect, $\mathbf m_i=\mathbf M_i/|\mathbf M_i|$ and $\gamma$ is the gyromagnetic ratio.  

In symmetric MTJ with the same FM metal in both electrodes the IEC effect is the even function of voltage, $J(V)=J(-V)$~[\onlinecite{Car2009,Swirkowicz2008,Butler2008,Butler2006,Brataas2008}]. However, from practical point of view the 
odd voltage dependence of the IEC is more useful. It would allow one to change the magnetic coupling type from FM 
to antiferromagnetic (AFM) and finally to realize controllable reversible magnetization switching in MTJ avoiding problems of dynamical SST-based remagnetization.

Theoretically it is shown that the odd contribution to $J(V)$ may appear for asymmetric MTJ (having different leads)~[\onlinecite{Car2009}]. IEC effect appears due to virtual hopping of electrons between FM leads and therefore is defined by the tunneling matrix. MTJ with different electrodes has asymmetric barrier resulting in the odd in voltage contribution to the tunneling probability. 
Weak odd contribution can also appear in the system with FM leads of different thickness~[\onlinecite{Stiles2008}]. 
Recently, the IEC was considered in asymmetric MTJ with ferroelectric (FE) barrier~[\onlinecite{Tsymbal2010}]. Following Ref.~[\onlinecite{Dai2016}] we will call such systems multiferroic tunnel junctions (MFTJ). In MFTJ the magnitude of 
IEC effect is defined by the direction of FE polarization. Switching of polarization direction changes the IEC strength. The dependence of the IEC effect on the polarization appears due to surface charges at the FE/FM interfaces in MTJ. They deform the barrier potential profile and therefore change tunneling matrix elements defining the IEC strength. These theoretical findings were not verified experimentally. IEC was  studied mostly in symmetric MTJs without FE (such as Fe/MgO/Fe).

New mechanism of IEC was recently proposed for MFTJ and granular multiferroics~[\onlinecite{Beloborodov2017ExGr,Beloborodov2017ExGr1,Beloborodov2017ExMTJ,Bel2014ME1,Bel2014ME2}]. 
The IEC may appear due to the spin-dependent part of the s-s electron Coulomb interaction. 
Two electrons in different leads experience the Coulomb based exchange interaction as in the Heitler-London model. Summation of the interaction over all electron pairs gives the magnetic coupling between leads. 
In the case of granular system the Coulomb blockade affects the IEC and also 
depends on dielectric constant. 
Interestingly, the IEC due to the many-body effects is inversely proportional to the FE barrier dielectric constant, $\epsi$.  In its turn $\epsi$ is voltage dependent in FEs and has the odd in voltage contribution leading to the odd component of the IEC effect in MFTJ.

In the present paper we will 
study one more mechanism leading to odd contribution in IEC voltage dependence in MFTJ. 
In contrast to our previous works~[\onlinecite{Beloborodov2017ExGr,Beloborodov2017ExGr1,Beloborodov2017ExMTJ,
Bel2014ME1,Bel2014ME2}], here we do not take into account the 
exchange interaction between s-s electrons and the Coulomb blockade. We focus on the hopping based IEC effect taking into 
account image forces acting on electrons in the barrier.
Image forces were neglected in previous studies of IEC effect. This is 
reasonable for MTJ without FE barrier since in this case the forces 
just reduce the barrier height independently of applied voltage. 
This is not the case in MFTJ. 
Image forces are inversely proportional to the barrier dielectric 
constant, $\epsi$. Since $\epsi$ is voltage dependent in FEs, the barrier 
reduction due to the image forces is also voltage dependent. This introduces an 
additional odd contribution to $J(V)$ dependence in MFTJ.

Recently, influence of image forces on electron transport in non-magnetic TJ with FE barrier and in granular metal with FE matrix was studied~[\onlinecite{Bel2014GFE2,Bel2014GFE1,Beloborodov2017TER}]. 
In the case of granular FE the image forces affect the Coulomb blockade effect leading to conductivity modification. In FE tunnel junctions the image forces essentially influence electron transport and cause significant electro-resistance effect. In the present manuscript we add magnetic degrees of freedom and study thermodynamic rather than transport properties of TJ.

The goals of this work are the following: 
1) study the influence of image forces on the IEC effect due to 
surface charges created by the FE polarization, Ref.~[\onlinecite{Tsymbal2010}]; 
2) investigate the ME effect appearing due to combination of image 
forces and voltage dependent dielectric constant 
of the FE barrier; 3) compare IEC voltage variation with STT effect.

TJ with FE barrier is currently not very well studied. Thin FE films lose their 
electrical properties as their thickness is reduced down to the nm scale range~[\onlinecite{Fridkin2006}].
Therefore, the fabrication technique for TJ with FE barrier is rather complicated. 
A lot of efforts were spent to investigate transport 
properties of non-magnetic TJ with FE barrier~[\onlinecite{Gruverman2013}]. The 
magneto-resistance effect in MFTJ was also studied~[\onlinecite{Li2011}]. 
Investigation of IEC effect in MFTJ is just started recently~[\onlinecite{Steren2016}].

The paper is organized as follows. 
The model of MTJ with FE barrier and calculations 
of IEC and STT effects are given in Sec. II.   In Sec. III subsections A and B 
we discuss general behavior of IEC and STT effects. 
Dependencies of IEC and STT effects on various parameters are discussed 
in subsections C-H of Sec. III. Last subsection of Sec. III is devoted 
to magnetization switching based on the IEC effect in MFTJ.

\section{The model}\label{Sec:AnFor}

In order to study the IEC in MFTJ we will use the following simple model. 
We assume that conduction electrons are responsible for the interlayer interaction. 
MFTJ consists of two homogeneously  magnetized FM layers and FE insulating spacer. Electrodes are thick enough and can be considered as infinite. A voltage $V$ is applied to the MFTJ. The 
FE barrier has polarization $P$, dielectric constant $\epsi$ and thickness $d$ (see Fig.~\ref{Fig:Model}). Polarization is uniform and directed along the x-axis. Both polarization $P$ and dielectric constant $\epsi$ are functions of the applied voltage (see details in Sec.~\ref{Sec:FEmodel}). The Hamiltonian for electrons is given by
\begin{equation}\label{Eq:LeadsHam1}
\hat H=\left\{\begin{split}
&-\frac{\hbar^2\Delta}{2m}-\Jsd(\hat \sigma \mathbf m_1)-U_\mathrm c+U_\mathrm p(x),~x<0,\\
&-\frac{\hbar^2\Delta}{2m_\mathrm e}+h_\mathrm b+U_\mathrm p(x)+U_\mathrm{if}(x)-eV\frac{x}{d},~0<x<d,\\
&-\frac{\hbar^2\Delta}{2m}-\Jsd(\hat \sigma \mathbf m_2)-U_\mathrm c+U_\mathrm p(x)-eV,~x>d,\\
\end{split}\right.
\end{equation}
where unit vectors $\mathbf m_{1,2}$ are directed along magnetizations of the leads, $\hat \sigma$ is the vector of Pauli matrices, $\Jsd$ is the spin subband splitting occurring due to interaction of conduction electrons with localized ones (as in Vonsovskii s-d model) or due to exchange interaction between conduction electrons themselves~[\onlinecite{Vons}], $x$ is the coordinate perpendicular to the layers surfaces, $m$ and $\me$ are the effective electron masses in metallic leads and the 
insulator layer. Generally the effective mass in FM leads may be spin dependent, especially in 
half-metals, but we do not take this into account for simplicity.  $h_\mathrm b$ defines the barrier height above the Fermi level of the left lead. We assume that the Fermi level of the left electrode corresponds to zero energy. The 
right lead is biased by the applied voltage. Quantities $-U_\mathrm c\pm J_\mathrm{sd}$ determine the bottoms of the conduction majority and minority spin bands. Using these quantities one can define the minority and majority Fermi momentums, $k^\pm_\mathrm{F}=\sqrt{2m(U_\mathrm c\pm J_\mathrm{sd})/\hbar^2}$. If $U_\mathrm c>\Jsd$ both minority and majority bands exist and we have two band ferromagnet (TBF). If $U_\mathrm c<\Jsd$, only one spin subband works and the leads are half-metal ferromagnets (HMF). The potential $U_\mathrm p$ appears due to the influence of the FE polarization and is defined as follows~[\onlinecite{Tsymbal2005}]
\begin{equation}\label{Eq:PolPot1}
U_\mathrm p=\phi_\mathrm p\left\{\begin{split}
&e^{x/\delta},~x<0,\\
&\left(1-\frac{2x}{d}\right),~0<x<d,\\
&-e^{-(x-d)/\delta},~x>d,\\
\end{split}\right.
\end{equation}
with the characteristic potential created by surface charges
\begin{equation}\label{Eq:PolPot2}
\phi_\mathrm p=\frac{edP\delta}{\eps0(d+2\epsi\delta)}.
\end{equation}
Here $e$ is the electron charge, $\eps0$ is the vacuum dielectric constant, $\delta$ is the Thomas-Fermi screening length. 
The potentials are found using electrostatic problem where FE with polarization $P$ is clamped in between two metals. The metals are treated within the Thomas-Fermi approximation with close circuit conditions.

The term $U_\mathrm{if}(x)$ in Eq.~(\ref{Eq:LeadsHam1}) describes the influence of the image forces.  They appear 
due to the interaction of electron inside the barrier and image charges occurring in metallic 
leads. We assume that electron concentration in the leads is high enough and the screening length in these metals $\delta$ is small enough ($\delta\ll d$). In this case one can use a simple picture of image forces inside the insulating barrier~[\onlinecite{Lundqvist}]
\begin{equation}\label{Eq:PotenIF}
U_\mathrm{if}(x)=\frac{0.795e^2d}{16\pi\eps0\epsi x(d-x)}.
\end{equation}
The terms with voltage $V$ in  Eq.~(\ref{Eq:LeadsHam1}) describe the effect of the applied voltage.
Following Refs.~[\onlinecite{Lundqvist},\onlinecite{Tsymbal2005}] we introduce the 
total potential barrier ``seen'' by tunneling electron as follows (region $0<x<d$)
\begin{equation}\label{Eq:BarrierGen}
U(x)=h_\mathrm b+U_\mathrm p(x)+U_\mathrm{if}(x)-eV\frac{x}{d},\\
\end{equation}

Calculating the image forces potential we treat the 
metallic leads as ideal neglecting corrections due to finite screening length. When calculating potentials $U_\mathrm p$ the finite screening length is crucial and can not be neglected. Note that $U_\mathrm{if}(x)$ diverges at points $x=0$ and $x=d$. In fact, 
in the vicinity of these points the image forces approximation does not work. The region where potential 
$U_\mathrm{if}(x)$ is not valid is defined by the size of the correlation hole size in metal. Usually it is 
of order of 0.05 nm. In this region the potential should smoothly transform from $U_\mathrm{if}(x)$ to the 
bulk metal potential. Since we consider FM lead the potential in the vicinity of metal/insulator interface should be spin dependent even inside the insulator. 
In our calculations we restricted image force potential 
at the level of the lower spin subband bottom. We tried a number of 
image forces potential shapes. All shapes give the same result, due to the fact that  
the shape of image forces potential in the close proximity to FM/FE interface 
influences only the low lying electron states. 
These electrons make small contribution to the overall IEC effect. 
The true potential profile in the vicinity of the interface is a long standing problem which is not fully resolved by now. Even \textit{ab initio} calculations based on density functional theory do not provide an
acceptable picture of the potential profile.

\subsection{FE layer}\label{Sec:FEmodel}

FE polarization below the Curie temperature is a function of
applied voltage and has a hysteresis.   We use the following formula to describe the 
FE polarization as a function of applied voltage
\begin{equation}\label{Eq:Pol}
P^{\pm}(V)=P_0 \frac{1-e^{-(V\mp V_\mathrm s)/\Delta V_\mathrm s}}{1+e^{-(V \mp V_\mathrm s)/\Delta V_\mathrm s}},
\end{equation}
where $V_\mathrm s$ is the switching voltage, 
$P_0$ is  the saturation polarization, $\Delta V_\mathrm s$ is the width of
the transition region. Superscripts ``$+$'' and ``$-$'' correspond to the upper and the
lower hysteresis branch, respectively.  For example, the polarization of HfZrO$_2$ is shown in Fig.~\ref{Fig:Model} and
can be approximately described with the following parameters:
$P_0=30$ $\mu$C/cm$^2$, $V_\mathrm s=d \cdot 10^8$ V (with $d$ being measured in m), and $\Delta V_\mathrm s=1 V_\mathrm s$. The parameters were obtained by fitting the experimental curves of Ref.~[\onlinecite{Mikolajick2012_1}].

The dependence of dielectric constant on voltage is given by the expression
\begin{equation}\label{Eq:Diel}
\epsi^{\pm}(V)=\varepsilon_\mathrm{min}+\frac{\Delta\varepsilon}{1+(V \mp V_\mathrm s)^2/\Delta V_\mathrm s^2}.
\end{equation}
This dependence captures the basic features of dielectric
constant behavior as a function of electric field.
The dielectric permittivity has two branches corresponding to two
polarization states. In the vicinity of the switching bias the dielectric permittivity,
$\epsi$ has a peak. For example, the dielectric constant of HfZrO$_2$ can be described using the
following parameters: $\varepsilon_\mathrm{min}=35$, $\Delta\varepsilon=15$ (see Fig.~\ref{Fig:Model}).

\subsection{Toy model of the electron potential profile}

On one hand the potential profile given in Eq.~(\ref{Eq:BarrierGen}) is complicated enough and does not allow for analytical solution of Shrodinger equation and calculating of wave functions. On the other hand the profile does not take into account several phenomena such as band structure of leads and the barrier, exchange-correlation effects in the vicinity of the 
FM/I interface. Since it does not allow making quantitative estimates of the IEC in MTJ we will 
further simplify the potential profile. Our simplification will not reduce the 
quantitative precision of our calculations but still capture the 
main physical phenomena that we will study in the present work.

The effect of the surface charges potential $U_\mathrm p$ on IEC is studied in Ref.~[\onlinecite{Vedyayev2005}]. Surface charges change the potential barrier for electrons and therefore change the tunneling probability. This effect can be captured within the quasiclassical approximation in which the average barrier height defines the tunneling probability. 
The influence of image force on the total barrier profile is twofold: 1) decreasing 
of the effective barrier width; and 2) decreasing of the effective barrier height. 
Generally, these two effects are also captured by the quasiclassical approximation.

We will change the initial barrier in Eq.~(\ref{Eq:BarrierGen}) with effective rectangular barrier of width $d_\mathrm{eff}$ and height $h_\mathrm{eff}$. Within the quasiclassical approximation the effective barrier thickness is defined by the intersection of the potential profile with the electron energy level. 
The points of intersection are $x_{1,2}$ and $d_\mathrm{eff}=x_2-x_1$. 
This point generally depends on the electron energy.

~

Note that due to the bias applied to the MFTJ the Fermi levels are different in the left and the right leads and the effective barrier ``seen'' by electrons in the 
right and the left electrodes is different. Generally, we can introduce a 
different barrier thickness for electrons in different leads $d_\mathrm{eff}^{\mathrm{r,l}}$. To 
determine $d_\mathrm{eff}^{\mathrm{l}}$ we find the 
intersection (points $x_{1,2}^\mathrm l$) of $U(x)$ with 
electron energy. To determine $d_\mathrm{eff}^{\mathrm{r}}$ we find intersections (points $x_{1,2}^\mathrm r$) of $U(x)$ with energy $-eV$ which is the Fermi level of the right lead). At zero bias $d_\mathrm{eff}^{\mathrm{r}}=d_\mathrm{eff}^{\mathrm{l}}$.

~

 If we neglect the image forces the effective barrier thickness is the same for electrons 
 in both leads and at any energy is equal to $d$. In the 
opposite situation when we neglect the effect of the bias and polarization the effective thickness 
at Fermi level is given by $d_\mathrm{eff}=d\sqrt{1-h_\mathrm c/h_\mathrm b}$.
The quantity $h_\mathrm c$ is given by 
\begin{equation}\label{Eq:ImFCharEn}
h_\mathrm c=\frac{0.795e^2}{4\pi\eps0\epsi d}.
\end{equation}
This is the characteristic potential associated with image forces in TJ. Particularly,  $h_\mathrm c$ is the reduction of the initial potential barrier height (see Eq.~(\ref{Eq:BarrierGen})) at the symmetry point ($x=d/2$) at zero bias.

Effective barrier height ``seen'' by electrons in different leads is also 
different and one can introduce $h_\mathrm{eff}^\mathrm{l,r}$. Even in the absence of the image forces and surface charges the applied bias leads to difference in the height of $eV$ and $h_\mathrm{eff}^\mathrm{r}=h_\mathrm{eff}^\mathrm{l}+eV$. 
A general expression for the effective barrier height has the form 
\begin{equation}\label{Eq:ImFCharEn}
\sqrt{h_\mathrm{eff}}=\frac{1}{d_\mathrm{eff}}\int_{x_1}^{x_2}\sqrt{U(x)-E}dx.
\end{equation}
Here $E$ is the electron energy counted from the Fermi 
level of the left electrode. Since due to the bias occupied 
electron energies are different in the left and the right lead one has different effective barriers.

In the next two subsections we provide simplified expressions 
for IEC and STT in MTJ. In these formulas we take into account 
only the electrons at the Fermi level and 
use $h_\mathrm{eff}^\mathrm{l,r}$ for effective barrier heights for electrons in the 
left and the right leads at their Fermi levels. 
While this approach is oversimplified and misses some important 
phenomena it still gives general trends of IEC and STT effect which is useful 
to keep in mind. In Sec.~\ref{Sec:III} we calculate IEC and STT effects taking into account all electrons at all energy levels (see Appendix~\ref{App:CalcDet}).

The most important phenomena that we study is the dependence of the IEC 
on dielectric properties of the barrier. The higher the dielectric constant the weaker the influence of image forces. For infinite $\epsi$ the image force potential disappears. This effect is captured in the toy potential. Similarly, the 
influence of voltage and FE polarization is also captured in this approach.

\subsection{Exchange interaction in MTJ}\label{Sec:ExMTJ}

Interlayer exchange coupling in MTJ can be described using the 
following macroscopic surface energy density, $-J(\mathbf m_1\cdot\mathbf m_2)$. IEC effect enters in to the Landau-Lifshitz-Gilber (LLG) equation as the torque,
$\dot{\mathbf m}_{1,2}=(\gamma J/(|\mathbf M_{1,2}| t_{1,2}))[\mathbf m_{1,2}\times \mathbf m_{2,1}]$. 
We will follow Slonczewski approach to calculate the IEC in MFTJ. The IEC is given by the equation
\begin{equation}\label{Eq:IEC_Gen}
J=-\sum_{i}Q^i_{y},
\end{equation} 
where $Q_y^i$ is the y spin component of the spin current density 
carried by the electron in the state $i$. The sum is over all orbital and spin states in both leads. 
Orbital state is described by quasimomentum inside each 
lead. Spin quantization axis for electrons in left (right) lead is along magnetization of left (right) electrode.
To calculate spin current of an electron in the state $i$ we find an electron wave function 
using the effective rectangular barrier model. We calculate the 
effective barrier height and width for each state $i$ 
(see more details in the Appendix~\ref{App:CalcDet}). The magnitude of the spin current 
depends on the mutual orientation of $\mathbf m_1$ and $\mathbf m_2$ as $\sin(\vartheta)$, 
where $\vartheta$ is the angle between $\mathbf m_1$ and $\mathbf m_2$. We calculate spin currents at $\vartheta=\pi/2$. 

Simplified analytical expressions for IEC in MFTJ can be found following Slonczewski approach~[\onlinecite{Slonczewski1989}]
\begin{equation}\label{Eq:Slonc1}
J=-\sum_{i=\mathrm{l,r}}\frac{\hbar^2(\varkappa_\mathrm{eff}^i)^4b^i}{4\me\pi^2(d^i_\mathrm{eff})^2 ((\varkappa_\mathrm{eff}^i)^2+(k_\mathrm F^+)^2)^2}e^{-2\varkappa_\mathrm{eff}^i d^\mathrm i_\mathrm{eff}},
\end{equation} 
where 
\begin{equation}\label{Eq:Slonc2}
\begin{split}
&b^i=
\frac{k_\mathrm F^+((k^+_\mathrm F)^2-\varkappa_\mathrm{eff}^i|k^-_\mathrm F|)}{(\varkappa_\mathrm{eff}^i+|k^-_\mathrm F|)},~\Jsd>U_\mathrm c,\\
&b^i=\frac{\varkappa_\mathrm{eff}^i ((\varkappa_\mathrm{eff}^i)^2-k^+_\mathrm F k^-_\mathrm F)(k^+_\mathrm F- k^-_\mathrm F)^2(k^+_\mathrm F+ k^-_\mathrm F)}{((\varkappa_\mathrm{eff}^i)^2+(k^-_\mathrm F)^2)^2},~\Jsd<U_\mathrm c.
\end{split}
\end{equation} 
Here we use the electron wave function inverse decay lengths, $\varkappa^{\mathrm{l,r}}_\mathrm{eff}=\sqrt{2\me h^{\mathrm{l,r}}_\mathrm{eff}/\hbar^2}$. The effective barrier parameters should be calculated at the Fermi levels of the 
left and the right electrodes, correspondingly. For zero voltage, $\varkappa^\mathrm{l}=\varkappa^\mathrm{r}$, and the 
above expression turns into Slonczewski formula. At finite bias the 
electron tunneling from the left electrode to the right one ``sees'' a different barrier 
comparing to electron moving in the opposite direction. 
This results in voltage dependence of IEC effect.

\subsection{Spin transfer torque}

The STT appears at finite bias~[\onlinecite{Slonczewski1989}]. 
This effect is described by a tensor with spin and orbital indexes. 
In our case the electron current flows along the x-axis. The 
tensor elements with orbital indexes $y$ and $z$ are zero. 
Therefore, we omit the orbital index in the spin current notation and keep only the 
spin index. The STT effect can not be associated with some energy contribution 
as the IEC effect~[\onlinecite{Slonczewski2005}]. It can be introduced 
into LLG equation for leads magnetizations as an 
additional torque in the form, 
$\dot{\mathbf m}_{1,2}=(\gamma J^{1,2}_\mathrm d/(|\mathrm M_{1,2}|t_{1,2}))[\mathbf m_{1,2}\times[\mathbf m_{1,2}\times \mathbf m_{2,1}]]$. 
Note that the magnitude of the torques acting on the left and the right leads 
can be different at finite voltage. System symmetry results in 
the following relation, $|J_\mathrm d^1(V)|=|J_\mathrm d^2(-V)|$. 
Below we will calculate the STT acting on the left lead and will omit the 
upper index. The STT effect does not exist without a charge current and 
is always related to energy dissipations. 
Therefore, we mark it with the subscript ``$\mathrm d$''.

STT has angular dependence similar to IEC, $\sin(\vartheta)$. 
For clarity we assume that $\mathbf m_1$ is along z-axis. Then, the
x-component of spin current flowing into the left lead is associated with STT effect. 
We neglect the z-component of the spin current when calculating the STT acting on 
the left lead. This is reasonable providing that the magnitude of magnetization is fixed by strong internal interaction. 
The STT effect  constant is given by the following expression 
\begin{equation}\label{Eq:STT_Gen}
J_\mathrm d=-\sum_{i}Q^i_x,
\end{equation}
where summation is only over electrons carrying the electric 
current (see Appendix~\ref{App:CalcDet} for details).

For analytical analysis one can use the simplified expression~[\onlinecite{Slonczewski1989}]
\begin{equation}\label{Eq:STT}
J_\mathrm d=\frac{eVe^{-2\varkappa_\mathrm{eff}d_\mathrm{eff}}}{2\pi^2d_\mathrm{eff}}D,
\end{equation}
where the quantity $D$ is given by 
\begin{equation}\label{Eq:STT2}
\begin{split}
&D=\frac{\varkappa_\mathrm{eff}^3(k_F^+)^2}{((\varkappa_\mathrm{eff})^2+(k_\mathrm F^+)^2)^2},~\Jsd>U_\mathrm c,\\
&D=\frac{\varkappa_\mathrm{eff}^3(\varkappa^4_\mathrm{eff}-(k_F^-)^2(k_F^+)^2)((k_F^+)^2-(k_F^-)^2)}{(\varkappa_\mathrm{eff}^2+(k_\mathrm F^-)^2)^2(\varkappa_\mathrm{eff}^2+(k_\mathrm F^+)^2)^2},~\Jsd<U_\mathrm c.
\end{split}
\end{equation}
The effective barrier parameter are calculated at the Fermi level
of the left (right) lead for positive (negative) $V$. 
Equation~(\ref{Eq:STT}) provides the linear in voltage term to the STT only. 
Therefore, it can not describe the STT voltage asymmetry 
known theoretically and experimentally~[\onlinecite{Butler2006,Brataas2008,Lee2009,SUZUKI2008,Ralph2008}]. 
In our numerical calculations we take this effect into account.

\section{Exchange interaction in MFTJ}\label{Sec:III}

In this section we will calculate the IEC and STT 
in MFTJ as a function of applied voltage depending on the system parameters.  

Voltage dependence of IEC can be considered as ME effect in MFTJ. 
In the literature a variation of the IEC constant with voltage 
is called field-like (or perpendicular) STT effect~[\onlinecite{Brataas2008,Lee2009,SUZUKI2008,Ralph2008}]. 
We will refer the IEC variation as the ME effect in the present work.

Variation of $J$ may cause a variation of the system magnetic state. Similarly, STT may also cause magnetization rotation.  STT and IEC effects produce spin currents flowing across the tunnel junction. 
The spin direction of these currents is different. The IEC produces the spin current perpendicular 
to magnetization plane while the STT causes the spin current in the plane of 
system magnetization. We will compare here only magnitudes of these spin currents $J$ and $J_\mathrm{d}$. 
Magnetization dynamics in MFTJ under the action of IEC variation and the STT effect  is beyond the
scope of the present manuscript and requires a separate investigation.

\subsection{System parameters}

Generally, the system has a lot of parameters. To reduce the number of variables we fixed some of them. 
We will use the parameters value chosen in this section in all cases below. 

On one hand  the FE can not be thinner than a single atomic layer, on the other hand the IEC itself decreases 
exponentially with $d$ for $d>0.5$ nm~[\onlinecite{Tsymbal2006,Lesnik2006,Schuhl2002}]. So, we will use $d=1$ nm in all our calculations keeping in mind that only for such a thin barrier the IEC has some impact on the 
magnetic state of MTJ.

In a similar way we will fix the barrier 
height $h_\mathrm b=0.5$ eV in all our calculations. Such a value 
is relevant for FE insulators. Obviously, increasing the height decreases the exchange interaction between the leads 
and the STT effect. Therefore, it is better to keep it as small as possible to be able to 
influence the magnetic state of MFTJ. 

Also we fix parameters of the FE polarization hysteresis loop 
such as switching voltage, $V_\mathrm s=0.1$ V  and the switching transition 
region, $\Delta V_\mathrm s=0.1$ V. 
They correspond to 1 nm thick layer of Hf$_{0.5}$Zr$_{0.5}$O$_2$ (see experimental 
data of Ref.~[\onlinecite{Mikolajick2012_1}]). 

We fix the effective mass in the barrier, $m_\mathrm e$ at the level of 0.4 electron mass. 
It follows from Eq.~(\ref{Eq:Slonc1}) that the IEC effect decreases with the barrier effective mass growth. In the absence of polarization and image forces the following relation holds $J(\alpha m_\mathrm e,h_\mathrm b)=(1/\alpha)J(m_\mathrm e, \alpha h_\mathrm b)$. The 
STT effect decreases slower than the IEC with the growth of barrier effective mass, 
$J_\mathrm d(\alpha m_\mathrm e,h_\mathrm b)=J_\mathrm d(m_\mathrm e, \alpha h_\mathrm b)$, see Eq.~(\ref{Eq:STT}).

Similar scaling rules can be written for effective mass of electron in FM leads, 
$J(\alpha m,U_\mathrm c, \Jsd)=J(m,\alpha U_\mathrm c, \alpha \Jsd)$ and $J_\mathrm d(\alpha m,U_\mathrm c, \Jsd)=J_\mathrm d(m,\alpha U_\mathrm c, \alpha \Jsd)$. In all calculations we use $m=0.9$ of electron mass. 

For good metals the parameter $\delta$ in  Eq.~(\ref{Eq:PolPot1}) is of 
order of 0.05 nm. Taking into account the fact 
that $\epsi\sim 50$  we have $(d+2\epsi\delta)\approx 2\delta\epsi$. In this regime $\delta$ 
vanishes in Eq.~(\ref{Eq:PolPot1}) and does not influence the IEC and STT effects. 
In our numerical calculations we fix the screening length, $\delta=0.05$ nm.

\subsection{General remarks}

In this section we use Eqs.~(\ref{Eq:BarrierGen}), (\ref{Eq:Slonc1}) and (\ref{Eq:STT}) to analyze general properties of IEC and STT effects.

1) Since we consider the symmetric MTJ the IEC effect does not 
depend on the direction of FE polarization 
at zero bias voltage. Spin transfer torque is absent at $V=0$.

2) The IEC depends on $(k_F^+-k_F^-)^2\sim \Jsd^2$ while the STT depends on $(k_F^+-k_F^-)^1\sim \Jsd^1$. This immediately leads to the fact 
that the STT becomes more important with decreasing of $\Jsd$.

3)  At zero polarization due to the system symmetry the IEC effect should be an even function of voltage. Neglecting the image charges one can get 
\begin{equation}\label{Eq:AvBarrierVolt}
\sqrt{h_\mathrm{eff}}\approx \sqrt{h_\mathrm b}\left(1-\frac{eV}{h_\mathrm b}-\frac{(eV)^2}{24h^2_\mathrm b}\right).
\end{equation} 
The linear term does not contribute to the IEC effect (which follows from the symmetry consideration). The quadratic term shows that effective barrier reduces with increasing voltage. Thus,
the IEC effect increases with voltage in agreement with other 
calculations~[\onlinecite{Butler2006}].

4) Neglecting the surface charges ($P_0=0$) and at $V=0$ one gets the following expression for the average barrier height at the Fermi level
\begin{equation}\label{Eq:AvBarrierImF}
\sqrt{h_\mathrm{eff}}\approx \sqrt{h_\mathrm b}\left(1-\frac{h_\mathrm c}{4h_\mathrm b}\mathrm{ln}\frac{h_\mathrm b}{4h_\mathrm c}\right).
\end{equation} 
One can see that the higher $\epsi$ the 
higher the barrier and the smaller $J$ and $J_\mathrm d$. This is the most general effect of the image forces. 

Note that many FEs (for example BTO or PZT) have very high dielectric 
constants ($\sim 1000$) making image forces negligibly weak. Image forces are significant in FEs with low dielectric constant only. 
There are a number of low dielectric constant FEs such as hafnium oxide family XHfO$_2$ (where X can be Y,  Co, Zr, Si)~[\onlinecite{Mikolajick2012,Hwang2014,Mikolajick2011}], rare-earth manganites XMnO$_3$ (where X is the rare-earth element)~[\onlinecite{Tokura2004}], colemanite~[\onlinecite{FATUZZO1960}], Li-doped ZnO~[\onlinecite{Liu2003}], etc. There are also numerous organic FEs with low dielectric constant~[\onlinecite{Boer2010,Xiong2012,Tokura2008}]. 

5) For voltage dependent dielectric constant the effective barrier 
height acquires an additional (to that shown in Eq.~(\ref{Eq:AvBarrierVolt})) dependence on voltage. If $\epsi (V)$ has the 
odd component (as it does in FE), the barrier height and the IEC would also have the 
linear in voltage contribution caused by image forces.

6) At zero voltage and neglecting image forces one can get the following estimate for the effective barrier
\begin{equation}\label{Eq:AvBarrierPol}
\begin{split}
\sqrt{h_\mathrm{eff}}&\approx \frac{1}{3\phi_\mathrm p}(\sqrt{h_\mathrm b+\phi_\mathrm p}^3-\sqrt{h_\mathrm b-\phi_\mathrm p}^3)\approx \\&\approx\sqrt{h_\mathrm b}\left(1-\frac{\phi^2_\mathrm p}{12h_\mathrm b^2}\right).
\end{split}
\end{equation} 
Surface charges reduce the barrier at zero voltage and increases the 
IEC and STT effects. This reduction of barrier appears due to internal 
electric field created by the surface charges. This effect is similar 
to the one discussed in the third clause.

At finite voltage the electric fields produced by 
polarization and voltage can be co-directed or counter-directed. For example, for positive voltage and positive polarization both electric 
fields are co-directed leading to the reduction of the barrier height and to the increase 
of the IEC effect. Negative voltage at $P>0$ decreases 
the interlayer coupling. 
FE polarization (surface charges) breaks the MTJ symmetry and results in the linear contribution to the 
voltage dependence of the IEC effect.

\subsection{IEC and STT as a function of spin subband splitting, $\Jsd$}

\begin{figure}
	\includegraphics[width=1\columnwidth]{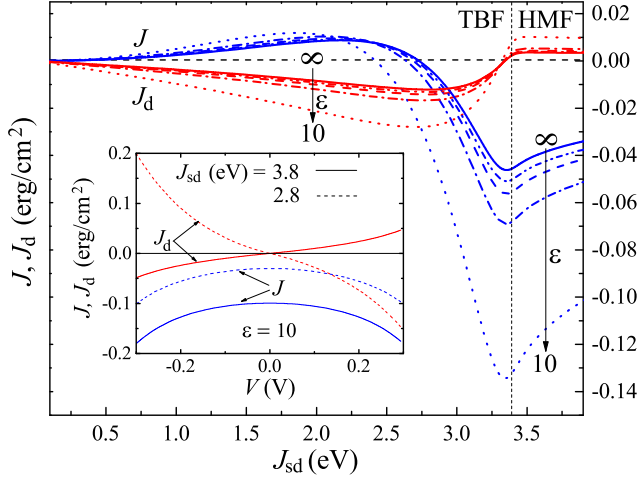}
	\caption{(Color online) The IEC ($J$) and STT ($J_\mathrm{d}$) effects as a function of the spin subband splitting $\Jsd$ at $V=0.1$ V and $U_\mathrm c=3.4$ eV (relevant for Co$_2$MnSi). Blue curves show $J$ and the red ones show $J_\mathrm{d}$. 
	The effect of FE polarization is neglected ($P_0=0$). Image forces are taken into account. Solid, dash-dot-dotted, dashed, dash-dotted and dotted curves correspond to $\epsi = \infty$, 100, 50, 25 and 10. Vertical line at $J_\mathrm{sd}=U_\mathrm c$ shows a border between two band ferromagnet and half-metal cases. Inset shows $J$ and $J_\mathrm{d}$ as a function of applied voltage $V$ for different $\Jsd$ at $U_\mathrm c=3.4$ eV and $\epsi=10$.}\label{Fig:JvsJsd}
\end{figure}

Figure~\ref{Fig:JvsJsd} shows a typical dependence of IEC and STT on 
the spin subband splitting, $J_\mathrm{sd}$, for the following parameters: 
$U_\mathrm c=3.4$ eV, $V=0.1$ V (STT is finite only for non-zero voltage). 
We neglect here the FE polarization ($P_0=0$) but consider the
image forces. We show the curves for several value of $\epsi$. The IEC is positive for small 
spin subband splitting and changes its sign for $\Jsd$ approaching $U_\mathrm c$. 
For HMF case ($\Jsd>U_\mathrm c$)  the IEC effect is negative. 
$J_\mathrm{d}$ has a similar behavior, changing its sign for large 
spin subband splitting.  

Note that spin subband splitting and Fermi energy can be tuned in many compounds varying 
proportions of material components. For example, the spin subband splitting in 
FM metal Co$_{1-x}$Fe$_x$S$_2$~[\onlinecite{Subramanian2004}] strongly depends on concentration of Fe and Co. At zero Fe concentration the material is a TBF. Increasing of Fe concentration transforms this material to a half-metal.

STT is much weaker for MFTJ with half-metal electrodes ($J_\mathrm{sd}>U_\mathrm c$),  comparing to the IEC constant $J$. Thus, in HMF region the voltage based variation of IEC (ME effect) is the main option for the control of MFTJ magnetic state. For two band FM metals both the 
STT and IEC variation are important. For 
small enough spin subband splitting the STT becomes the most important mechanism causing MTJ magnetic dynamics under applied voltage.

Parameters $J$ and $J_\mathrm{d}$ depend on dielectric constant of the barrier $\epsi$. Increasing $\epsi$ increases the average barrier height 
according to our estimates, Eq.~(\ref{Eq:AvBarrierImF}). This leads to decreasing of coupling and the 
STT effect. One can see that variation of $J$ with $\epsi$ is significant and comparable to the 
value of the STT effect.

\subsection{Dependence of the IEC and STT on voltage in MTJ without FE}

Inset in Fig.~\ref{Fig:JvsJsd} shows the IEC and STT effects as a function of applied voltage in MTJ with simple insulating barrier ($P_0=0$ and $\epsi$ is voltage independent). We use  $\epsi=10$ as in MgO barrier. Two curves for IEC correspond to different values of spin 
subband splitting, $J_\mathrm{sd}$. It is known~[\onlinecite{Car2009}] that 
variation of IEC with applied voltage is comparable to the STT in MTJ. 
STT is asymmetric function (see red lines) 
of voltage in agreement with previous theoretical and experimental 
studies~[\onlinecite{Butler2006,Brataas2008,Lee2009,SUZUKI2008,Ralph2008}]. 
Dependence of IEC on voltage is an even function of $V$ due to symmetry of the tunnel 
junction without FE barrier. 

\begin{figure}
	\includegraphics[width=1\columnwidth]{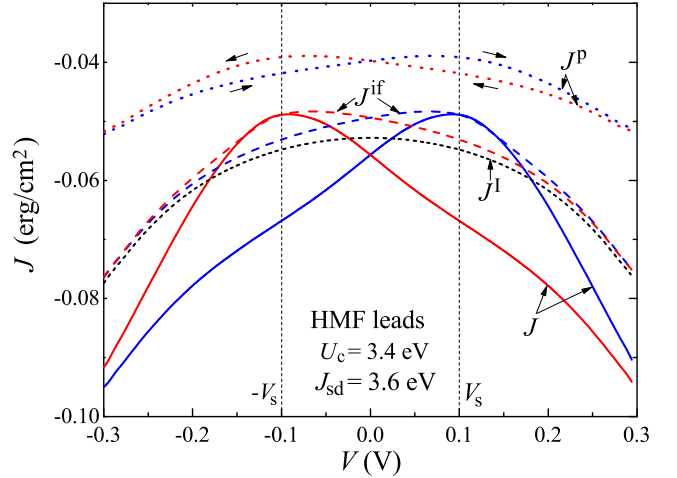}
	\caption{(Color online) IEC ($J$) for MTJ with FE barrier as a function of voltage $V$. Leads parameters are $\Jsd=3.6$ eV and $U_\mathrm c=3.4$ eV (relevant for Co$_2$MnSi HMF). Dotted lines show IEC neglecting image forces but taking into account FE surface charges, $J^\mathrm p$. $P(V)$ is given by Eq.~(\ref{Eq:Pol}) with $P_0=30~\mu$C/cm$^2$ (relevant for HfZrO$_2$ FE). Blue and red curves correspond to different hysteresis branches. Dashed lines show IEC taking into account only image forces, $J^\mathrm{if}$. Voltage dependence of dielectric constant $\epsi$ is given by Eq.~(\ref{Eq:Diel}) with $\varepsilon_\mathrm{min}=30$ and $\Delta \epsi=15$ (relevant for HfZrO$_2$ FE). Solid lines are the IEC taking both effects into account, $J$. Black short-dashed line shows IEC for MTJ with simple insulating barrier with no polarization and voltage-independent dielectric constant $\epsi=30$, $J^\mathrm I$. Arrows show hysteresis bypass direction.}\label{Fig:JvsV_IFvsP}
\end{figure}

\subsection{Influence of image forces and surface charges on the IEC in MFTJ}

Both the surface charges (and associated potential $U_\mathrm p$) and the image forces ($U_\mathrm{if}$) influence the IEC effect in MTJ with FE barrier. To study their effect we calculate three different quantities $J^\mathrm{if}$, $J^\mathrm{p}$ and $J$. 
The first one, $J^\mathrm{if}$, is the IEC effect taking into account only image forces and 
neglecting the potential $U_\mathrm p$ (we put $P_0=0$ but use the 
voltage-dependent $\epsi$ in Eq.~(\ref{Eq:Diel})).  $J^\mathrm{p}$ is calculated neglecting $U_\mathrm{if}$ (image forces) but taking into account surface charges described by $U_\mathrm p$. Quantity $J$ accounts for both effects. 

Figure~\ref{Fig:JvsV_IFvsP} shows the IEC effect as a function of voltage for the case of HMF leads with $U_\mathrm c=3.4$ eV and $\Jsd=3.6$ eV. These parameters correspond to HMF Co$_2$MnSi~[\onlinecite{Dederichs2006,Kubota2006}]. The HMF has a 
rather complicated band structure. We model it with free electron model. Majority and minority spin bands 
bottoms are taken from \textit{ab initio} calculations. Effective mass is found by fitting the density of states at the Fermi level in the majority spin band to the density of states in \textit{ab initio} calculations. The chosen effective mass ($m=0.9$ of free electron mass) corresponds to HMF Co$_2$MnSi. Note that this material demonstrates half-metallic properties and high tunneling magneto-resistance in MTJ structure~[\onlinecite{Kubota2006, Yamamoto2007}] in contrast to many HMFs losing their high spin polarization at an interface~[\onlinecite{Groot2008}]. Curves in Fig.~\ref{Fig:JvsV_IFvsP} correspond to FE barrier with saturation polarization $P_0=30~\mu$C/cm$^2$. 
Variation of dielectric constant in the barrier is described by Eq.~(\ref{Eq:Diel}) with $\varepsilon_\mathrm{min}=30$ and $\Delta\epsi=15$. These parameters correspond to Hf$_{0.5}$Zr$_{0.5}$O$_2$ FE~[\onlinecite{Mikolajick2012}]. 

Black short dashed line in Fig.~\ref{Fig:JvsV_IFvsP} shows 
the IEC ($J^\mathrm I$) as a function of voltage neglecting the dependence of 
polarization ($P_0=0$) and dielectric constant ($\epsi=30$) on bias.  
Nevertheless the image forces are taken into account. This corresponds to the case of a 
simple insulator. The dependence is an even function of voltage. 
Variation of IEC with voltage is significant (about 25\% in the shown 
voltage range). Increasing of IEC magnitude with voltage is caused by 
reducing of the effective barrier when bias voltage is applied (see Eq.~(\ref{Eq:AvBarrierVolt})).

Dotted line in Fig.~\ref{Fig:JvsV_IFvsP} shows the 
IEC in MFTJ neglecting image forces, $J^\mathrm{p}$. The exchange coupling as a 
function of voltage has two branches corresponding to two different FE polarization 
states, $P^+$ and $P^-$.  Arrows indicate the path of the hysteresis loop. 

Two maxima at $V=\pm V_\mathrm s$ correspond to polarization switching. 
As we stated the electric fields induced by polarization and bias voltage can be 
either co-directed or counter-directed. Switching between these two 
situations happens at $V=\pm V_\mathrm s$. This explains the occurrence of these two maxima.

At zero voltage the IEC has the same value for both branches due 
to the system symmetry. At finite voltage the symmetry is broken and the 
IEC depends on polarization state. At small voltages one can write, 
$J^\mathrm p\sim J_0^\mathrm p+\alpha PV$. This is in contrast to the situation 
considered in Ref.~[\onlinecite{Tsymbal2005}] where MTJ with different magnetic 
leads show a different IEC effect at zero voltage for two different polarization states.  

Note that even though the image forces are not taken into account when calculating $J^\mathrm p$ the voltage dependence of dielectric constant, $\epsi$ still influences the IEC. According to Eq.~(\ref{Eq:PolPot1}) the 
dielectric constant defines the potential barrier disturbance by the surface charges. The higher $\epsi$ the lower the 
influence of surface charges. The influence of 
$\epsi$ bias dependence on surface charge potential and 
on the tunneling electro-resistance and magneto-resistance was considered in Ref.~[\onlinecite{Xiao2012}]. 
Similarly, $\epsi(V)$ contribute to the IEC variation in the present model.

Influence of the surface charges on the IEC effect does not exceed several percent which is less than IEC variation without FE barrier (black short dashed line).

Influence of image forces alone is shown with red and blue dashed lines, $J^\mathrm{if}$. Behavior of $J^\mathrm{if}$ is similar to $J^\mathrm{p}$. It has two branches corresponding to two branches of dielectric constant $\epsi^+$ and $\epsi^-$, Eq.~(\ref{Eq:Diel}). At zero bias the 
IEC is the same for both branches. According to our estimates (Eq.~(\ref{Eq:AvBarrierImF})) 
the increase of dielectric permittivity leads to the decrease 
of tunneling probability and therefore to the decrease 
of exchange coupling. The dielectric constant has maxima 
at $V=\pm V_\mathrm s$. This explains two peaks of IEC at $V=\pm V_\mathrm s$.

The change of IEC effect due to image forces alone is large 
than due to surface charges, but still not as large 
as changes caused by voltage itself. We estimate the 
IEC changes due to image forces $\Delta J^\mathrm{if}$ as difference between IEC for the 
upper (or lower) branch at $V=V_\mathrm s$ and at $V=0$ ($\Delta J^\mathrm{if}=J^\mathrm{if}(V_\mathrm s)-J^\mathrm{if}(0)$). IEC variation due to voltage can be estimated as difference between $J^\mathrm{I}$  at finite and at zero voltage
 ($\Delta J^\mathrm{I}=J^\mathrm{I}(V)-J^\mathrm{I}(0)$). One can see that at $V=V_\mathrm s$ the 
 quantity $\Delta J^\mathrm{if}$ is larger 
 than $\Delta J^\mathrm{I}$, but with increasing voltage $\Delta J^\mathrm{I}$ exceeds the 
 IEC changes due to image forces.
 \begin{figure}
	\includegraphics[width=1\columnwidth]{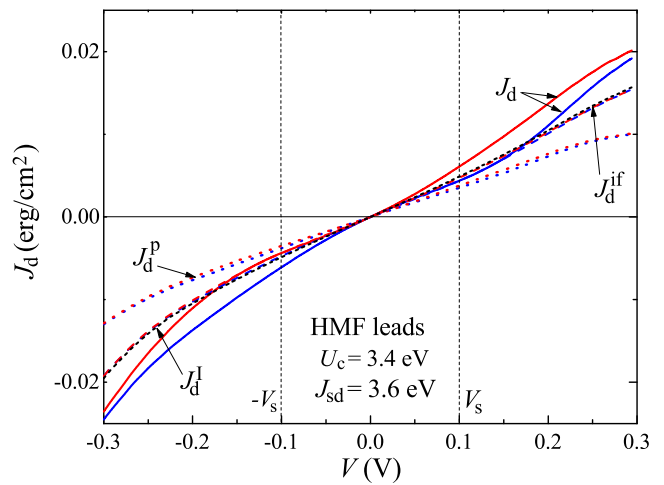}
	\caption{(Color online) STT ($J_\mathrm d$) for MTJ with FE barrier as a function of voltage $V$. All parameters are the same as in the previous figure. Dotted lines show IEC neglecting image forces but taking into account FE surface charges, $J^\mathrm p_\mathrm d$.  Dashed lines show IEC taking into account only image forces, $J^\mathrm{if}_\mathrm d$. Solid lines are the IEC taking both effects into account, $J_\mathrm d$. Blue and red curves correspond to different hysteresis branches. Black short-dashed line shows IEC for MTJ with simple insulating barrier, $J^\mathrm I_\mathrm d$ with no polarization and voltage-independent dielectric constant.}\label{Fig:STTvsV_IFvsP}
\end{figure}

The situation changes drastically, when both image forces and surface charges are taken into account. 
Corresponding curves ($J(V)$) are shown with red and blue solid lines in  Fig.~\ref{Fig:JvsV_IFvsP}. The curves have similar shape as $ J^\mathrm{if}$ and $ J^\mathrm{p}$, but have much larger variation. Moreover, the IEC has strong linear dependence at voltages $|V|<V_\mathrm s$, 
which may be useful for applications (see Sec.~\ref{Sec:Appl}). Thus, the curves demonstrate that 
image forces together with surface charges essentially change the dependence of the IEC effect on applied voltage in MFTJ.

\subsection{Influence of image forces and surface charges on STT in MFTJ}

Image forces and surface charges influence the STT effect in MFTJ as well. However, their influence in this case is not very pronounced. Figure~\ref{Fig:STTvsV_IFvsP} shows the voltage dependence of the STT magnitude. The same parameters and 
notations are used as in the previous figure. Superscript ``$\mathrm I$'' means that 
the STT is calculated for MTJ with an insulator barrier ($P_0=0$ and $\epsi=30$ is  voltage independent. Superscript ``$\mathrm{if}$'' 
stands for STT calculated in the presence of image forces and voltage dependent dielectric constant, 
but for $P_0=0$. $J^\mathrm p$ stands for STT effect accounting for surface charges in the absence of 
image forces. The STT in the presence of both effects is denoted with $J_\mathrm d$. 
Neither image forces nor surface charges qualitatively change the 
STT voltage behavior. However, a weak hysteresis appears when both effect are taken into account. 
\begin{figure}
	\includegraphics[width=1\columnwidth]{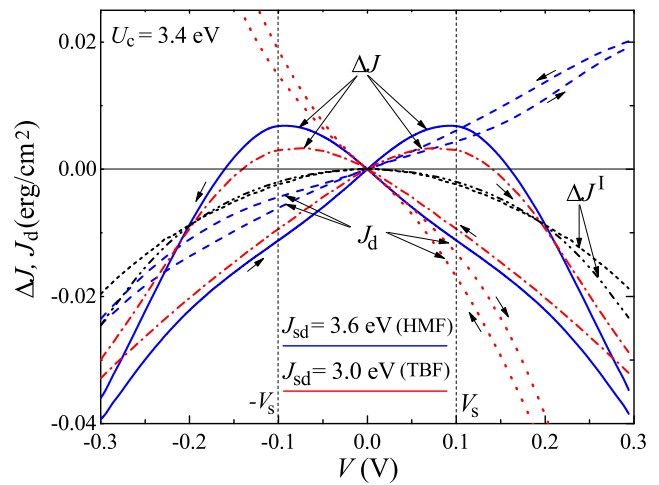}
	\caption{(Color online) STT ($J_\mathrm d$) and IEC variation ($\Delta J$) for MFTJ as a function of voltage $V$. 
	We use $U_\mathrm c=3.4$ eV, $P_0=30~\mu$C/cm$^2$, $\varepsilon_\mathrm{min}=30$, $\Delta\epsi=15$. Solid and dash-dotted lines show IEC variation ($\Delta J$) at $\Jsd=3.6$ eV (HMF case) and $3.0$ eV (TBF case), correspondingly.  Dashed and dotted lines 
	are the STT effect ($J_\mathrm d$) at $\Jsd=3.6$ eV and $3.0$ eV, correspondingly. All curves demonstrate 
	hysteresis loop. Arrows indicate the hysteresis loop bypass direction. 
	Black short dashed and dash dot dotted lines show IEC for non-FE insulating barrier 
	with $\epsi=30$  at $\Jsd=3.6$ eV and $3.0$ eV, correspondingly.}\label{Fig:STTvsV_IFvsP}
\end{figure}

\begin{figure*}[t]
	\begin{center}
		\includegraphics[width=1\textwidth, keepaspectratio]{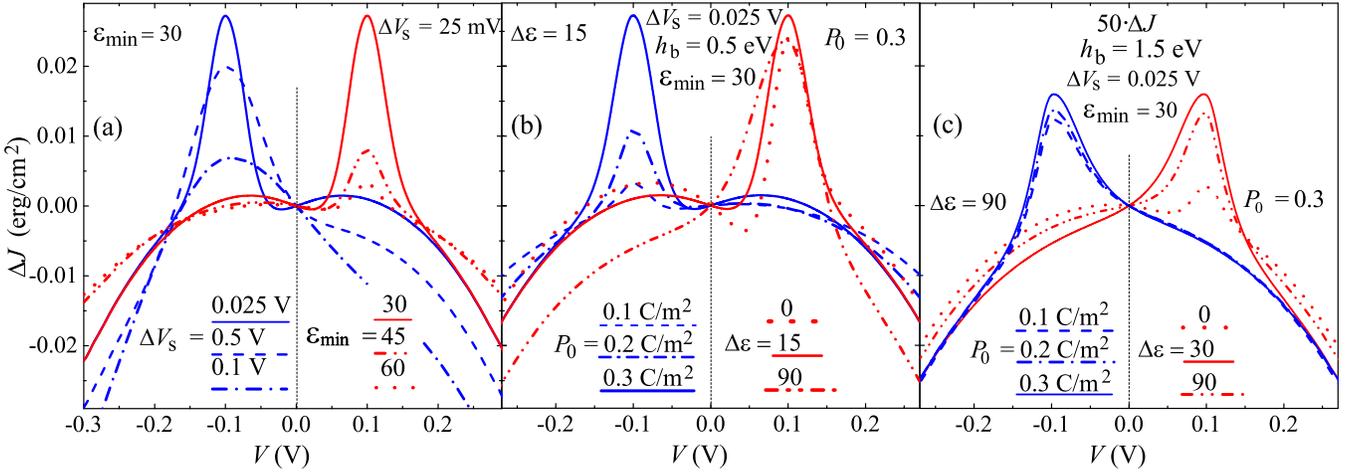}
		\caption{(Color online) ME effect ($\Delta J (V)$) for various barrier parameters. (a) Blue lines show upper branches of $\Delta J(V)$ for different polarization switching transition region $\Delta V_\mathrm s$ at $\epsi_\mathrm{min}=30$. Red lines are the lower branches of $\Delta J(V)$ for different minimum dielectric constant $\epsi_\mathrm{min}$ at $\Delta V_\mathrm s=25$ mV. Other parameters are the following: $h_\mathrm b=0.5$ eV, $\Delta \epsi=15$, $P_0=30~\mu$C/cm$^2$. (b) Blue lines show upper branches of $\Delta J(V)$ for different saturation polarization $P_0$ at $\Delta \epsi=15$. Red lines are the lower branches of $\Delta J(V)$ for different dielectric constant variation values $\Delta \epsi$ at $P_0=30~\mu$C/cm$^2$. Other parameters are the following: $h_\mathrm b=0.5$ eV, $\epsi_\mathrm{min}=30$, $\Delta V_\mathrm s=25$ mV. 
		(c) Value of $\Delta J$ is multiplied by 50. Blue lines show upper branches of $\Delta J(V)$ for different saturation polarization $P_0$ at $\Delta \epsi=90$. Red lines are the lower branches of $\Delta J(V)$ for different dielectric constant variation values $\Delta \epsi$ at $P_0=30~\mu$C/cm$^2$. Other parameters are the following: $h_\mathrm b=1.5$ eV, $\epsi_\mathrm{min}=30$, $\Delta V_\mathrm s=25$ mV. For all plots we use $U_\mathrm c = 3.4$ eV, $\Jsd=3.6$ eV. }
		\label{Fig:JvsV_PvsEps}
	\end{center}
\end{figure*}

\subsection{STT vs IEC in MFTJ}

In this section we compare the magnitude of STT effect and 
variation of IEC effect with voltage (see Fig.~\ref{Fig:STTvsV_IFvsP}). 
We subtract the IEC effect at zero voltage from the $J(V)$ dependence 
introducing the notation, $\Delta J(V)=J(V)-J(0)$. Here we take 
both image forces and surface charges into account. Two cases are shown: $\Jsd>U_\mathrm c$ (HMF case) and 
$\Jsd<U_\mathrm c$ (TBF case). In the case of two band FM leads the STT effect 
grows fast with voltage and exceeds the IEC variation (compare red dotted and red dash 
dotted lines in Fig.~\ref{Fig:STTvsV_IFvsP}). Thus, mostly the STT is 
responsible for magnetization dynamics in this case. In the case of HMF 
leads the IEC variation becomes stronger than the STT effect 
(compare blue dashed and blue solid lines in Fig.~\ref{Fig:STTvsV_IFvsP}). 
In this case mainly the IEC defines the magnetization dynamics and even the 
magnetic state of MFTJ. 

Black dash dot dotted and short dashed lines ($\Delta J^\mathrm I$) shows the 
IEC variation in MTJ without FE barrier. Coupling changes caused by image forces 
and surface charges essentially exceed the ones caused by the 
voltage itself ($\Delta J^\mathrm I$), at least in the region of voltages 
below the FE switching ($|V|<V_\mathrm s$). 

Thus, image forces and surface 
charges result in enforcement of IEC variation with voltage. Such variations 
become stronger than the STT effect in MFTJ with HMF leads. It is 
even more important that the image forces and surface charges produce the 
linear in voltage contribution to the IEC effect, which does not occur in 
symmetric MTJ without FE barrier.

\subsection{Variation of IEC as a function of barrier parameters}

Figure~\ref{Fig:JvsV_PvsEps} shows dependence of IEC variation $\Delta J$ on 
various barrier parameters. Here we calculate the IEC by taking into account 
both surface charges and image forces. Panel (a) shows 
$\Delta J$ as a function of minimum dielectric constant, $\epsi_\mathrm{min}$ 
and polarization switching region width, $\Delta V_\mathrm s$. 
Blue lines show $\Delta J(V)$ for three different $\Delta V_\mathrm s$ for
fixed $\epsi_\mathrm{min}=30$. These curves correspond to the upper branch of 
FE hysteresis loop. The dependence for lower branches is a mirror reflection 
with respect to zero voltage, $V=0$. Red curves show $\Delta J(V)$ for tree 
different $\epsi_\mathrm{min}$ at fixed $\Delta V_\mathrm s=0.025$ V. Only lower 
hysteresis branches are shown. All curves are for the following parameters: 
$h_\mathrm b=0.5$ V, $\Delta \epsi=15$, $P_0=30~\mu$C/cm$^2$, $U_\mathrm c=3.4$ eV, $\Jsd=3.6$ eV. 

The peak of IEC variation at $V=-V_\mathrm s$ grows with 
decreasing of the transition region width $\Delta V_\mathrm s$. 
This can be understood as follows. According to 
Eqs.~(\ref{Eq:AvBarrierImF}) and (\ref{Eq:AvBarrierPol}) the decrease of dielectric constant 
decreases the effective barrier due to both surface charges and 
image forces. The decrease of transition region, $\Delta V_\mathrm s$ leads to the reduction 
of $\epsi$ at zero voltage (see Eq.~(\ref{Eq:Diel})). The dielectric constant 
at $V=-V_\mathrm s$ stays the same, $\epsi(-V_\mathrm s)=\epsi_\mathrm{min}+\Delta\epsi$. 
Finally, the IEC effect grows at $V=0$ and stays the same at $V=-V_\mathrm{s}$ leading to peak growth. 

Red curves show that the IEC variation strongly depends on the minimum 
dielectric constant, $\epsi_\mathrm{min}$. Growth of $\epsi_\mathrm{min}$ leads 
to the decrease of IEC variation. The strength of image forces and potentials 
created by the surface charges are inversely proportional to the 
dielectric constant. Therefore, increasing of minimum value 
of $\epsi$ reduces the effect of surface charges and image forces. 

Panels (b) and (c) provide an additional insight into the effect of 
image forces on IEC. Both panels show $\Delta J(V)$ for different 
values of saturation polarization, $P_0$ and variation of dielectric constant, 
$\Delta \epsi$. Blue (red) curves show modification of $\Delta J(V)$ with 
varying of $P_0$ ($\Delta \epsi$). We use the following 
parameters:  $U_\mathrm c=3.4$ eV, $\Jsd=3.6$ eV and $\Delta V_\mathrm s=0.025$ V. 
In panel (b) the curves correspond to the low barrier system with $h_\mathrm b=0.5$ eV. 
In this case $P_0$ (see blue curves) influences the IEC effect much stronger than 
the $\Delta \epsi$ (see red curves). We conclude that in this case the IEC variation 
appears due to modulation of FE polarization rather than due to the 
modulation of dielectric constant with voltage. However, image forces essentially 
enhance the effect of surface charges. Panel (b) shows the case of high 
barrier, $h_\mathrm b=1.5$ eV. In this case situation is the 
opposite: $\Delta J(V)$ curves weakly depend on polarization 
(see blue curves) and strongly depend on $\Delta\epsi$ (see red curves). 
This means that the dependence of dielectric constant on voltage is the main 
source of the IEC variation (ME effect) and one can neglect the variation of polarization. 

To conclude, the role of image forces is twofold: 1) they enhance the 
IEC variations caused by surface charges, even in the absence of {voltage dependence of} $\epsi$ and 
2) due to variation of dielectric constant with voltage the image forces cause the 
ME effect which can be even stronger than that caused by surface charges.

\subsection{Magnetization switching in the MFTJ}\label{Sec:Appl}

\begin{figure}
	\includegraphics[width=1\columnwidth]{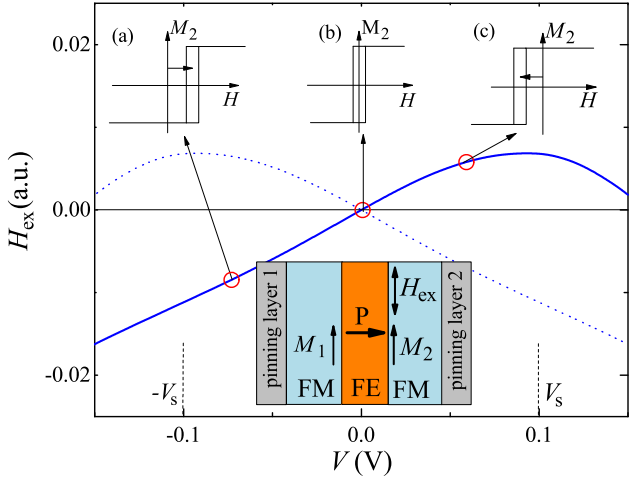}
	\caption{(Color online) MFTJ with two pinning layers. Left pinning layer creates very strong exchange bias 
	acting on magnetization $M_1$. IEC effect between layers $M_1$ and $M_2$ creates exchange field $H_\mathrm{ex}$ acting on the ``free'' layer $M_2$. Right pinning layer compensates the exchange field at zero voltage. Therefore at zero voltage ``free'' magnetic layer has symmetric hysteresis loop (marked with (b)). 
	Positive voltage shifts the hysteresis loop of ``free'' layer to the left as shown in (c). 
	Negative voltage moves the hysteresis loop right, (a). 
	Swiping voltage one can switch magnetization of the ``free'' layer. }\label{Fig:MagnSwitch}
\end{figure}

Dependence of IEC on voltage can be used for switching of magnetization in MFTJ. 
Importantly, the switching mechanism is essentially different comparing to the 
STT effect. The STT effect may be used for dynamical switching only. 
It does not correspond to any energy contribution in the system Hamiltonian 
and can cause rotation of magnetization independently of magnetization state. 
In contrast, the IEC effect defines the system ground state. As it was shown 
above the interlayer coupling may have a linear contribution as a function 
of voltage. At zero voltage the IEC is finite. Consider the following system. 
The left FM layer is pinned (see Fig.~\ref{Fig:MagnSwitch}) 
and can not be switched. At zero bias this layer produces an 
effective field acting on the right layer $H_\mathrm{ex}(V=0)=J(V=0)/(t_2 M_2)$. The 
right layer has small enough coercive field and is pinned also.
The second pinning layer is chosen such that it compensates the IEC effect at zero voltage. 
Pinning magnetic field acting on the right layer is $H_\mathrm b=-H_\mathrm{ex}(V=0)$. 
In this case the sign of the effective IEC (IEC effect + pinning) depends on voltage sign. 
Positive voltage would create the FM IEC while the negative voltage 
creates the AFM IEC. If coercive field of the right layer is smaller than the IEC variation, 
one can switch the magnetization with voltage. There
is no need to tune voltage impulse parameters to get reliable switching, 
in contrast to STT based remagnetization. The IEC variation of $\Delta J=0.02$ erg/cm$^2$ 
creates an effective field $H_\mathrm{ex}(V_\mathrm s)-H_\mathrm{ex}(0)=80$ Oe acting 
on the material with saturation magnetization $M_\mathrm s=500$ Gs and thickness $t=5$ nm. Therefore, if the 
material has a coercive field lower than 80 Oe, one can switch it with electric field.

For example, half-metal F$_3$O$_4$ has a rather small magnetization at 
room temperature about 140 Gs, meaning that the effective IEC field 
variation can be as high as 250 Oe which is the same as 
the coercive field of the material~[\onlinecite{Zhou2001}].

HMF considered in the present work Co$_2$MnSi has the 
coercive field in the range of 10 to 100  Oe depending 
on fabrication conditions~[\onlinecite{Zabel2010,Harris2001}]. 
Thus, it can be also switched with IEC variation effect.

\section{Conclusion}

We studied voltage dependence of the IEC and STT effects in MTJ with FE barrier. 
We took into account two phenomena influencing the IEC in MFTJ: 1) modification of the tunnel barrier potential due to the FE polarization; 2) modification of the barrier due to image forces acting on electrons in the tunneling spacer. 
Voltage dependent 
polarization results in the voltage dependent IEC effect. The 
influence of the image forces is twofold: i) they enhance the IEC variation 
occurring due to FE polarization and ii) image forces in combination with 
voltage dependent dielectric constant of the FE barrier produce an 
additional contribution to the interlayer coupling variation. This contribution becomes 
the dominant for MFTJ with high barrier. We compare the IEC with 
the STT effect occurring at finite voltage in MTJ. In the case of HMF the 
variation of IEC effect is the dominant mechanism defining MTJ magnetic 
state and dynamics. We estimate the variation of IEC effect and show that 
it can be used for controllable switching of magnetization in MFTJ with HMF leads.

\section{Acknowledgements}
This research was supported by NSF under Cooperative Agreement Award EEC-1160504 
and NSF PREM Award. O.U. was supported by Russian Science Foundation (Grant  16-12-10340).

\appendix

\section{IEC and STT formulas}\label{App:CalcDet}

Consider two FM leads separated by a rectangular barrier of width $d$ ($0<x<d$). 
Magnetizations of the leads are orthogonal. Left (right) lead magnetization is along 
the z(x)-axis. Consider a majority spin electron with unit incident flux in the 
left lead with energy $E$ moving to the barrier. The quasimomentum of up (or majority) 
band electron in the left lead is $\klp$. For down (minority) spin band the 
quasimomentum is $\klm$ in the left lead. In the right lead the 
quasimomentum of up (down) spin electron with energy $E$ is $\krp$ ($\krm$). 
Inside the barrier the electron wave function decays. The barrier potential 
is spin-independent. Therefore, the decay length, $\varkappa$ is spin-independent,
however it is energy dependent. Lets calculate the spin current of a 
single electron inside the barrier. Spin current flows along the x-axis 
(across the barrier). We omit the spatial direction index in the notation 
of spin current and keep only the spin index.
We consider several different cases depending on electron energy, bias voltage and 
spin subband splitting.

First consider the case when all $\klp$, $\klm$, $\krp$ and $\krm$ are real. 
In this case we have
\begin{equation}
Q^{+}_x=\frac{4\hbar^2}{m_\mathrm e}\frac{\klp\varkappa^2(\krp-\krm)(\varkappa^2-\krp\krm)e^{-2\varkappa d}}{(\varkappa^2+(\klp)^2)(\varkappa^2+(\krp)^2)(\varkappa^2+(\krm)^2)},
\end{equation}
\begin{equation}
Q^{+}_y=-\frac{4\hbar^2}{m_\mathrm e}\frac{\klp\varkappa^3((\krp)^2-(\krm)^2)e^{-2\varkappa d}}{(\varkappa^2+(\klp)^2)(\varkappa^2+(\krp)^2)(\varkappa^2+(\krm)^2)},
\end{equation}
\begin{equation}
Q^{+}_z=\frac{4\hbar^2}{m_\mathrm e}\frac{\klp\varkappa^2(\krp+\krm)(\varkappa^2+\krp\krm)e^{-2\varkappa d}}{(\varkappa^2+(\klp)^2)(\varkappa^2+(\krp)^2)(\varkappa^2+(\krm)^2)}.
\end{equation}
These expressions coincide with Slonczewski formulas if one put $\klp=\krp$ and $\klm=\krm$.
The spin current for electron incident on the barrier from the left 
electrode with down spin can be found using the following relations $Q^{-}_{x,z}=-Q^{+}_{x,z}(\klp\leftrightarrow\klm,\krp\leftrightarrow\krm)$, $Q^{-}_{y}=Q^{+}_{y}(\klp\leftrightarrow\klm,\krp\leftrightarrow\krm)$.

Note that here we use slightly different 
notations comparing to the main text (see Eq.~(\ref{Eq:IEC_Gen})). The 
upper index here stands for spin state of an electron only, 
while in the main text the index was responsible for both spin and orbital states.

Now consider the case when only the majority spin states in both 
leads are allowed and the minority spin states can not 
propagate ( $\klp$ and  $\krp$ are real, $\klm$ and $\krm$ are imaginary).
\begin{equation}
Q^{+}_{x,z}=\frac{4\hbar^2}{m_\mathrm e}\frac{\klp\krp\varkappa^2e^{-2\varkappa d}}{(\varkappa^2+(\klp)^2)(\varkappa^2+(\krp)^2)},
\end{equation}
\begin{equation}
Q^{+}_y=\frac{4\hbar^2}{m_\mathrm e}\frac{\klp\varkappa^2(\varkappa|\krm|-(\krp)^2)e^{-2\varkappa d}}{(\varkappa^2+(\klp)^2)(\varkappa^2+(\krp)^2)(\varkappa+|\krm|)}.
\end{equation}
Minority (down) spin states do not propagate in this 
case and $Q_{x,y,z}^{-}=0$. This result agrees with 
Slonczewski result for one band leads (if one put $\klp=\krp$ and $\klm=\krm$).

However, at finite voltage (or in MTJ with different leads) 
the situations can occur when only one spin channel is ``active''
in one reservoir and two spin channels are available in the other. 
Consider the case when both spin bands are active in the 
left electrode and only one works in the right lead.
\begin{equation}
Q^{+}_{x,z}=\frac{4\hbar^2}{m_\mathrm e}\frac{\klp\krp\varkappa^2 e^{-2\varkappa d}}{(\varkappa^2+(\klp)^2)(\varkappa^2+(\krp)^2)},
\end{equation}
\begin{equation}
Q^{+}_y=\frac{4\hbar^2}{m_\mathrm e}\frac{\klp\varkappa^2(\varkappa|\krm|-(\krp)^2)e^{-2\varkappa d}}{(\varkappa^2+(\klp)^2)(\varkappa^2+(\krp)^2)(\varkappa+|\krm|)}.
\end{equation}  
Expression for electron in minority spin band can be obtained with 
the substitutions $Q^{-}_{x,z}=-Q^{+}_{x,z}(k^+_\mathrm l\to k^-_\mathrm l)$, $Q^{-}_{y}=Q^{+}_{y}(k^+_\mathrm l\to k^-_\mathrm l)$.

In the opposite case when only one spin band 
is ``active'' in the left electrode and two spin bands are available in the right electrode we have
\begin{equation}
Q^{+}_x=\frac{4\hbar^2}{m_\mathrm e}\frac{\klp\varkappa^2(\krp-\krm)(\varkappa^2-\krp\krm)e^{-2\varkappa d}}{(\varkappa^2+(\klp)^2)(\varkappa^2+(\krp)^2)(\varkappa^2+(\krm)^2)},
\end{equation}
\begin{equation}
Q^{+}_y=-\frac{4\hbar^2}{m_\mathrm e}\frac{\klp\varkappa^3((\krp)^2-(\krm)^2)e^{-2\varkappa d}}{(\varkappa^2+(\klp)^2)(\varkappa^2+(\krp)^2)(\varkappa^2+(\krm)^2)},
\end{equation}
\begin{equation}
Q^{+}_z=\frac{4\hbar^2}{m_\mathrm e}\frac{\klp\varkappa^2(\krp+\krm)(\varkappa^2+\krp\krm)e^{-2\varkappa d}}{(\varkappa^2+(\klp)^2)(\varkappa^2+(\krp)^2)(\varkappa^2+(\krm)^2)}.
\end{equation}
The formulas are the same as for the two band case (considered first),
but down spin electron are absent in the left lead in this case and $Q^{-}_{x,y,z}=0$. 

The last case is related to the situation 
when there are no states in both spin bands in the right electrode. 
The right electrode behaves as an insulator in this energy region. 
In this case there is no electron flow across the 
barrier and $Q^{{+},{-}}_{x,z}=0$. The y-component is not zero 
\begin{equation}
Q^{+}_y=-\frac{4\hbar^2}{m_\mathrm e}\frac{\klp\varkappa^2(|\krp|-|\krm|)e^{-2\varkappa d}}{(\varkappa^2+(\klp)^2)(\varkappa+|\krp|)(\varkappa+|\krm|)},
\end{equation}
\begin{equation}
Q^{-}_y=\frac{4\hbar^2}{m_\mathrm e}\frac{\klm\varkappa^2(|\krp|-|\krm|)e^{-2\varkappa d}}{(\varkappa^2+(\klm)^2)(\varkappa+|\krp|)(\varkappa+|\krm|)}.
\end{equation}

Spin current created by an electron incident on the barrier from the 
right electrode can be found by the following relations   $Q_x|_{\mathrm{right}}=-Q_z|_{\mathrm{left}}(k^\pm_\mathrm{l}\leftrightarrow k^\pm_\mathrm{r})$. 
The sign ``$-$'' occurs because the electron moves in the 
opposite direction in comparison to the electron in the 
left electrode. 
At the same time $Q_y|_{\mathrm{right}}=Q_y|_{\mathrm{left}}(k^\pm_\mathrm{l}\leftrightarrow k^\pm_\mathrm{r})$.

The IEC effect is related to the y-component of the spin current. 
To calculate the constant $J$ one has to sum the 
spin current $Q_y$ over all spin and orbital states in both leads. 
To calculate 
the y-component of the spin current created by the left lead electrons we use the equation
\begin{equation}
Q^\mathrm l_y=\frac{1}{8\pi^2}\sum_i\int_0^{k^i_\mathrm{F}}dk k ((k^i_{F})^2-k^2) Q^i_y.
\end{equation}
To simplify formulas we omit the factor $\hbar^2/2m$ below. 
In the above equation we use in $Q^{+}_y$ the following quasimomentums $k^+_\mathrm{l}=k$, $k^-_\mathrm{l}=\sqrt{k^2-2J_\mathrm{sd}}$, $k^+_\mathrm{r}=\sqrt{k^2-eV}$, $k^+_\mathrm{r}=\sqrt{k^2-(eV+2J_\mathrm{sd})}$. Spin current $Q^{-}_y$ should be used with $k^+_\mathrm{l}=\sqrt{k^2+2J_\mathrm{sd}}$, $k^-_\mathrm{l}=k$, $k^+_\mathrm{r}=\sqrt{k^2-(-2J_\mathrm{sd}+eV)}$ and $k^-_\mathrm{r}=\sqrt{k^2-eV}$. For each energy $E=k^2$ we calculate the 
effective barrier thickness $d_\mathrm{eff}$ and the 
inverse decay length, $\varkappa_\mathrm{eff}$. The effective barrier 
thickness is defined by intersection of 
potential Eq.~(\ref{Eq:BarrierGen}) with energy level $E=k^2$. 
At the same energy we determine the effective barrier height.   

Spin current created by the right electrode $Q^\mathrm{r}_y$ is given by the same equation but we introduce $k^+_\mathrm{r}=k$, $k^-_\mathrm{r}=\sqrt{k^2-2J_\mathrm{sd}}$, $k^+_\mathrm{l}=\sqrt{k^2+eV}$, $k^-_\mathrm{l}=\sqrt{k^2+(eV-J_\mathrm{sd})}$ for majority spin channel and  $k^+_\mathrm{r}=\sqrt{k^2+2J_\mathrm{sd}}$, $k^-_\mathrm{r}=k$, $k^+_\mathrm{l}=\sqrt{k^2+(2J_\mathrm{sd}+eV)}$ and $k^-_\mathrm{l}=\sqrt{k^2+eV}$ for minority spin channel. 
Effective barrier thickness and height are calculated in the same manner as for the left electrode.

Total y-component of spin current is given by $Q^\mathrm{tot}_y=Q^\mathrm l_y+Q^\mathrm r_y$.

To calculate the x component of the spin current we 
sum only over the electron producing the charge current. Thus, we have for $V>0$ 
\begin{equation}
Q^\mathrm{tot}_x=\frac{1}{8\pi^2}\sum_i\int_0^{k^i_\mathrm{F}}dk k F^i(k) Q^i_x,
\end{equation}
where 
\begin{equation}
F^i(k)=\left\{\begin{split}
&V,~ (k^\mathrm{i}_\mathrm{F})^2-k^2>eV,\\
&(k^\mathrm{i}_\mathrm{F})^2-k^2,~ \mathrm{overwise}.
\end{split}\right.
\end{equation}
We use the same quasimomentums as in the case of y-component of 
spin current produced by electrons in the left electrode. 
For $V>0$ the right electrode does not contribute to the x-component 
of the spin current. For negative bias $V<0$ we have 
\begin{equation}
Q^\mathrm{tot}_x=\frac{1}{8\pi^2}\sum_i\int_0^{k^i_\mathrm{F}}dk k F^i(k) Q^i_z.
\end{equation}
Note that for negative bias only the right electrode contribute to the 
x component of the spin current. Therefore, we introduce 
the same quasimomentums as we use for calculating of y-component 
of spin current carrying by electrons in the right lead 
but with $V\to -V$. Note, that to calculate the x-component of the 
spin current at negative bias we use $Q_z$. This is due 
to relations introduced above.

\bibliography{FTJ}

\end{document}